%% file: Project.tex
\documentclass[review,final]{elsarticle}

\journal{Journal of Sound and Vibration}
\usepackage[mathscr]{eucal}
\usepackage{amsmath}
\usepackage{amssymb}
\usepackage{amsthm}
\usepackage{graphicx} 
\usepackage{epsfig}
\usepackage{ifthen}
\usepackage{psfrag}
\usepackage[utf8]{inputenc}
\usepackage[english,russian]{babel}
\usepackage[T2A]{fontenc}
\usepackage{setspace}
\usepackage{misccorr}
\usepackage{url}
\usepackage[colorlinks,allcolors=blue,unicode,pdftitle={Passage through resonance}]{hyperref}
\usepackage{stackrel}
\usepackage{mathtools}

\input def


\begin{document}
\selectlanguage{english}
\begin{frontmatter}
\title{
Passage through a resonance for a mechanical system, having time-varying parameters and
possessing a single trapped mode. The principal term of the resonant solution}
\author[ipme]{E.V.~Shishkina}
\ead{shishkina\_k@mail.ru}
\author[ipme,spbstu]{S.N.~Gavrilov\corref{mycorrespondingauthor}} 
\ead{serge@pdmi.ras.ru}
\cortext[mycorrespondingauthor]{Corresponding author}
\author[ipme]{Yu.A.~Mochalova}
\ead{yumochalova@yandex.ru}
\address[ipme]{Institute for Problems in Mechanical Engineering RAS, V.O., Bolshoy
pr. 61, St.~Petersburg, 199178, Russia}
\address[spbstu]{Peter the Great St.~Petersburg Polytechnic University (SPbPU),
Polytechnicheskaya str.~29, St.Petersburg, 195251, Russia}


\selectlanguage{english}



\input def-beam


\input{passage}

\end{document}

%% file: def.tex
\def\TODO#1{\marginpar{\setstretch{1}\vspace*{0mm}\hrule\strut\vphantom{p}\scriptsize #1\strut\vphantom{p}\hrule}}

\def\d{\mathrm d}
\def\defe{\buildrel{\text{def}}\over=}
\def\tp{t^\prime}  
\def\prpr#1{#1^{\prime\prime}}
\def\thet#1{(\ref{#1})}
\def\veps{\varepsilon}
\def\Int{\int_{-\infty}^{+\infty}}
\def\IInt{\iint_{-\infty}^{+\infty}}
\def\VP{\operatorname {Vp}}
\def\PV{\operatorname {Vp}}
\def\const{\operatorname {const}}
\def\sign{\operatorname {sign}}
\def\Sqrt{\sqrt{\omega^2+k}}

\def\SQRTO{\sqrt{1-\Omega_0^2}}

\def\erf{\operatorname{erf}}
\def\S{{S}}
\def\C{{C}}
\def\defe{\buildrel{\text{def}}\over=}
\def\KK{{k_0}}
\def\Farg{\frac{\varkappa}{\sqrt{2\pi\mu}}}
\def\sign{\operatorname{sign}}
\def\const{\operatorname{const}}
\def\opm{{\omega^\pm}}
\def\Iint{\iint_{-\infty}^{+\infty}}  
\def\Np{\mathbf N^\prime}
\def\Rp{\mathbf R^\prime}
\def\bbU{\mathbf U}
\def\bbE{\mathbf E}
\def\bbF{\mathbf F}
\def\bbf{\mathbf f}
\def\bbN{\mathbf N}
\def\bbG{\mathbf G}
\def\bbL{\mathbf L}
\def\bbR{\mathbf R}
\def\bmR{\bmathcal R}
\def\bmF{\bmathcal F}
\def\mF{\mathcal F}
\def\bbr{\mathbf r}
\def\bbK{\mathbf K}
\def\bbi{\mathbf i}
\def\bbj{\mathbf j}
\def\ii{\bbi\,{\otimes}\,\bbi}
\def\dxi#1{{#1}_\xi}
\def\ddt#1{{#1}_{\tau\tau}}
\def\dt#1{{#1}_{\tau}}
\def\ddxit#1{{#1}_{\xi\tau}}
\def\ddxi#1{{#1}_{\xi\xi}}
\def\cD{\mathcal D}
\def\cK{\mathcal K}
\def\cEps{\mathcal E}
\def\pd#1#2{\dfrac{\partial#1}{\partial#2}}
\def\W{\mathscr W_0}

\def\equ{equation\ }
\def\equs{equations\ }

\def\Re{\operatorname{Re}}
\def\Im{\operatorname{Im}}
\def\I{\mathrm i}
\def\NEW#1{{\color{black} #1}}

\def\erf{\operatorname{erf}}
\def\aaa{\Omega_{01}}

\newenvironment{NEWe}{\color{black}}{}

%% file: def-beam.tex
\def\defe{\buildrel{\text{def}}\over=}
\def\tp{t^\prime}  
\def\prpr#1{#1^{\prime\prime}}
\def\thet#1{(\ref{#1})}
\def\veps{\varepsilon}
\def\Int{\int_{-\infty}^{+\infty}}
\def\IInt{\iint_{-\infty}^{+\infty}}
\def\VP{\operatorname {Vp}}
\def\PV{\operatorname {Vp}}
\def\const{\operatorname {const}}
\def\sign{\operatorname {sign}}
\def\Sqrt{\sqrt{\omega^2+k}}
\def\erf{\operatorname{erf}}
\def\S{{S}}
\def\C{{C}}
\def\defe{\buildrel{\text{def}}\over=}
\def\KK{{k_0}}
\def\Farg{\frac{\varkappa}{\sqrt{2\pi\mu}}}
\def\sign{\operatorname{sign}}
\def\const{\operatorname{const}}
\def\opm{{\omega^\pm}}
\def\Iint{\iint_{-\infty}^{+\infty}}  
\def\Np{\mathbf T^\prime}
\def\Rp{\mathbf R^\prime}
\def\bbU{\mathbf U}
\def\bbE{\mathbf E}
\def\bbF{\mathbf F}
\def\bbf{\mathbf f}
\def\bbN{\mathbf T}
\def\bbG{\mathbf G}
\def\bbL{\mathbf L}
\def\bbR{\mathbf R}
\def\bmR{\boldsymbol{\mathscr R}}
\def\bmF{\boldsymbol{\mathscr F}}
\def\mF{\mathscr F}
\def\bbr{\mathbf r}
\def\bbK{\mathbf K}
\def\bbi{\mathbf i}
\def\bbj{\mathbf j}
\def\ii{\bbi\,{\otimes}\,\bbi}
\def\dxi#1{{#1}_\xi}
\def\ddt#1{{#1}_{\tau\tau}}
\def\dt#1{{#1}_{\tau}}
\def\ddxit#1{{#1}_{\xi\tau}}
\def\ddxi#1{{#1}_{\xi\xi}}
\def\cD{\mathscr D}
\def\cK{\mathscr K}
\def\cEps{\mathscr E}
\def\pd#1#2{\dfrac{\partial#1}{\partial#2}}

\def\w{{u}}
\def\u{{w}}

\def\equ{Eq.~}
\def\equs{Eqs.~}

\theoremstyle{remark}
\newtheorem*{Remark}{\it Remark}
\newtheorem{remark}{\it Remark}

\def\TODO#1{\marginparwidth=15mm
\marginpar{\hrule\strut\vphantom{p}\scriptsize #1\strut\vphantom{p}\hrule}}

\def\varGamma{\mathscr D}

\def\GG{\mathscr G}

%% file: passage.tex
%
\begin{abstract}
We consider a forced oscillation and passage through resonance for an
infinite-length system, having time-varying parameters and possessing a single
trapped mode.  The system is a string, lying on the Winkler foundation and
equipped with a discrete linear mass-spring oscillator of time-varying
stiffness. We obtain the principal term of the asymptotic expansion for the
resonant solution describing the motion of the inclusion (i.e., the
mass-spring oscillator). The obtained result was verified by independent
numerical calculations based on solution of the corresponding partial
differential equation by means of
the method of finite differences. The comparison demonstrates a good
agreement in a neighbourhood of the instant of resonance. 
\end{abstract}


\end{frontmatter}

\section{Introduction}
In the paper, 
we consider a forced oscillation of an infinite-length
mechanical system having time-varying parameters and possessing a single trapped
mode \cite{gavrilov-da70}
characterized by a frequency $\Omega_0(\epsilon \tau)$ (here and in what follows,
$\epsilon$ is a small parameter, $\tau$ is the time). 
The system is a string, lying on the Winkler foundation, and equipped with a
discrete linear mass-spring oscillator of time-varying stiffness. 
In the case of a constant spring stiffness, the spectrum of natural
oscillations for such a mechanical system may contain unique (positive) eigenvalue, which is less than the
lowest frequency for the string on the uniform foundation (the cut-off
frequency). This special natural frequency
corresponds to a trapped mode of oscillation with eigenform localized near
the spring. The phenomenon of
trapped modes was discovered in the theory of surface water waves
\cite{ursell1951trapping}.

The discrete
oscillator is subjected to a harmonic external force with constant frequency
$\hat\Omega$. In the case of the passage through the resonance, we obtain the
principal term of the asymptotic expansion describing the motion of the
inclusion (i.e., the mass-spring oscillator). To do this, we use a
combination of two asymptotic approaches. 

The first approach was suggested in
\cite{gavrilov2002etm} to  
describe a {\it free} localized oscillation in systems possessing a single
trapped mode.
It was successfully used to investigate various free localized oscillation of
spatially non-uniform infinite length systems with time-varying properties 
\cite{Gavrilov-2006-trans,indeitsev2016evolution,gavrilov2016trapped,
gavrilov2017trapped,shishkina2018non,Gavrilov_2019,gavrilov-da70}. 
In particular, in our recent study \cite{gavrilov-da70} we used this approach to
study a free localized oscillation in the system considered in this paper.
The extensive
bibliography on trapped modes and localized waves in infinite-length linear 
systems can be found in 
recent papers \cite{shishkina2018non,Gavrilov_2019} and monograph
\cite{Ind-book-R2E}.
The second approach was used in
\cite{kevorkian1971passage,kevorkian1974erratum}
to describe a {\it forced} oscillation and passage through a resonance in a
single degree of freedom system (a linear oscillator). 

The paper by Fowler \& Lock \cite{Fowler1922} was probably the first one
(see \cite{nayfehperturbation}), where the resonant excitation of a linear
system of ODE \NEW{(ordinary differential equations)} with slowly varying coefficients was considered from the
asymptotic point of view. In the series of studies by 
Feschenko, Shkil \& Nikolenko summarized in monograph
\cite{feschenko1967eng}, the authors obtained an asymptotically simplified system
of ODE by describing the passage through a resonance in a system with several degrees
of freedom (the same approach is discussed in \cite{nayfehperturbation}).
Kevorkian in \cite{kevorkian1971passage} obtained, using the method of multiple
scales, the matched asymptotic
expansion that is uniformly valid both in the resonant case (``the inner
expansion'') and in the non-resonant case (before and after the resonance, ``the
outer expansions''). In \cite{Ablowitz1973}
Ablowitz, Funk \& Newel demonstrated that the Kevorkian's solution for times
greater than the instant of resonance contained an error: the 
oscillation of order $1/\sqrt{\epsilon}$ 
emerged near the resonance should never
vanish after the resonance. This fact is in contradiction with results of
\cite{kevorkian1971passage}. 
The error was also pointed out in \cite{Gautesen1974}, where the
problem was solved by a modified WKB \NEW{(the Wentzel--Kramers--Brillouin)} approach. 
The error was resolved in
erratum \cite{kevorkian1974erratum}. Skinner 
\cite{Skinner1997} applied the stationary phase method to obtain a uniformly
valid solution of the linear problem. Modification of the Kevorkian
procedure for the case of a weakly nonlinear system with slowly varying
properties are considered in many studies, see e.g.
\cite{Ablowitz1973, Kevorkian1974, Lewin1978, Kevorkian1980, Kevorkian1980a}
and many other references.

\begin{NEWe}   
Our work have at least two important motivations.
First, we can discover trapped modes in many systems used
as simple analytical models for engineering applications. For instance, an
infinite continuous system with discrete inertial moving loads under certain
conditions can possess trapped modes 
\cite{gavrilov2002etm,kaplunov1986torsional,Vesnitskii1992,Vesnitskii1996,indeitsev2000resonance,Sergeev2005,Roy2018},
though this fact is not always properly recognized in the corresponding
studies. The trapped mode frequency is a resonance frequency for the infinite
mechanical system.
It is not very difficult to imagine a
real engineering problem, where it is necessary to consider the passage through
resonance for such a system.
For example, such a problem naturally emerges if we want to consider
a non-uniformly moving discrete oscillator subjected to a harmonic force (some
kind of extension of the problem considered in recent paper \cite{Roy2018}).

The second motivation is a more theoretical one. Let us imagine a mechanical
system with several identical distant discrete inclusions and corresponding
trapped modes with frequencies close to each other \cite{Glushkov2011a}. We
can naturally transform such a system to a system with slowly time-varying
parameters if we additionally assume that inclusions can slowly move along the
continuous system. In the latter case, we obtain a linear system with internal
resonances of a fundamentally new type (the resonances between different
weakly coupled trapped modes), which has not been considered, as far as we know,
in the literature
before. The problem considered in this paper is our very initial attempt to
describe such resonances analytically.\footnote{\NEW{Note that our initial interest was
related with an unfinished (for the time being) attempt to take into account
the existence of a trapped mode \cite{indeitsev2000resonance} in analytical
treatment of the problem for a semi-infinite system with a single moving inclusion
considered in \cite{Vesnitskii1992,Vesnitskii1996}. The problem for the latter
semi-infinite system, in fact, can be equivalently transformed to a problem
for an infinite system with two inclusions and two coupled trapped
modes.}}  Accordingly, we have chosen a model problem (a string equipped with an oscillator of time-varying stiffness) 
to prepare the mathematical tricks, which we plan to use in future for more
complicated and better motivated (by engineering or physical practice) problems.

With regard to the following problems related to 
\begin{enumerate} 
\item free oscillation for a linear oscillator with time-varying
stiffness;
\item 
resonant forced oscillation for a linear oscillator with time-varying
stiffness;
\item free oscillation for a system, having time-varying
parameters and possessing a single trapped mode;
\item resonant forced oscillation for the latter system;
\end{enumerate}
we claim that they are
quite different from the asymptotic point of view.
For problems 1 \& 2 this difference is
well-known (see e.g. \cite{kevorkian1971passage,kevorkian1974erratum,nayfehperturbation,feschenko1967eng}). 
In our previous studies 
\cite{gavrilov-da70,gavrilov2002etm,Gavrilov-2006-trans,indeitsev2016evolution,gavrilov2016trapped,gavrilov2017trapped,shishkina2018non,Gavrilov_2019}
we have
demonstrated that problems of the third type are much more difficult and complicated than problem~1. 
And now, in the current manuscript, we deal with problem~4. To the best of our
knowledge, this paper is the first study, where passage through a resonance
for such 
a system is considered.

%
\end{NEWe}

The paper is organized as follows.
In Section~\ref{PAS-SEC:formulation} we consider the formulation of the
problem. 
In Section~\ref{PAS-SEC:known} we present a summary of some known
results related with free oscillation in the system under consideration. These
results are necessary to consider the forced oscillation. In Section~\ref{PAS-SEC:pas}
the forced oscillation and passage through the resonance are considered. We
obtain the principal singular term of the ``inner expansion''. In the
particular case of a discrete oscillator with sufficiently large stiffness and mass,
the obtained formulas transform into formulas obtained in 
\cite{kevorkian1971passage,kevorkian1974erratum}. The particular cases of an 
increasing and a decreasing stiffness are considered in
Sections~\ref{PAS-SS1} and \ref{PAS-SS2}, respectively. 
In Section~\ref{PAS-section-numerics} we verify the obtained analytical
results by independent numerical calculations
based on solution of the corresponding \NEW{partial differential equation} by means of the method
of finite differences. The comparison of the analytical and numerical
solutions demonstrates a good agreement
in a neighbourhood of the instant of resonance.
In Conclusion (Section~\ref{PAS-SEC:Concl}) we discuss the basic results of
the paper, future plans and possible applications.

\section{Mathematical formulation}
\label{PAS-SEC:formulation}
The governing equations for the system in a dimensionless form are
\begin{gather}
\prpr u-\ddot u
-u=-P(\tau)\,\delta(\xi),
\label{PAS-maineq-SPRING}
\\
P(\tau)=
-M\ddot u(0,\tau)-K(\epsilon\tau)u(0,\tau)+p(\tau).
\label{PAS-force-spr}
\end{gather}
Here, and in what follows, we denote by prime the derivative
with respect to the spatial coordinate $\xi$ and
by overdot the derivative with respect to
the time $\tau$,
$u(\xi,\tau)$ is the displacements,
$p(\tau)$ is the given external force on the discrete oscillator,
$P(\tau)$ is the unknown force on the string from the discrete oscillator,
constant $M\geq0$ is the  mass in the discrete oscillator,
$K(\epsilon\tau)$ is the 
spring stiffness for the discrete oscillator (a given slowly varying function
of time), 
$\epsilon$ is a formal small parameter.
We do not assume that $K>0$ (hence, the spring stiffness
can be negative 
\cite{gavrilov2016trapped,shishkina2018non,gavrilov2017trapped,
Gavrilov_2019}
or zero).
The initial conditions for Eq.~\eqref{PAS-maineq-SPRING} can be
formulated in the following form
\cite{Vladimirov1971}:
\begin{equation}
u \big|_{\tau<0}\equiv0.
\label{PAS-<0-SPRING}
\end{equation}
In order to consider forced oscillation, we put
\begin{equation}
p(\tau)=
H({\tau})\exp{(-\I\hat{\Omega}\tau)},
\label{PAS-external-force}
\end{equation}
where $\hat\Omega=\const$ is the resonant frequency, $H({\tau})$ is the Heaviside function. 

The problem under consideration 
\eqref{PAS-maineq-SPRING},
\eqref{PAS-force-spr},
\eqref{PAS-<0-SPRING}
is symmetric with respect to $\xi=0$. 
{Integrating 
\eqref{PAS-maineq-SPRING}
over $\xi=0$
results in the following condition:}%
\footnote{It is only the highest spatial derivative term in the left-hand side 
of Eq.~\eqref{PAS-maineq-SPRING} that can involve
a delta-like term \cite{Vladimirov1971}}
\begin{equation}
\big[u'(0,\tau)\big]=-{P(\tau)}=
M\ddot u(0,\tau)+K(\epsilon\tau)u(0,\tau)-p(\tau).
\label{PAS-1dof-SPRING-p}
\end{equation}
Here, and in what follows, $[\mu]\equiv\mu(\xi+0)-\mu(\xi-0)$ for any arbitrary
quantity~$\mu$. Expression 
\eqref{PAS-1dof-SPRING-p} is the equation for balance of momentum formulated
for the mass-spring oscillator attached to the string. In fact, this is the
Rankine-Hugoniot jump condition 
\cite{Courant1999}
formulated at the fixed position $\xi=0$.
It also can be obtained from
the first principles (see, e.g.,
\cite{Gavrilov(ActaMech),gavrilov2016revisitation,Ferretti_2019}, where the more complicated case of
geometrically nonlinear model of the string with a point moving load is considered). 
Due to the symmetry, one has $[u']=2u'(\xi+0)$. Thus, the problem for the
infinite string can be equivalently reformulated as the problem for the
homogeneous equation 
\begin{gather}
\prpr u-\ddot u
-u=0\qquad (\xi>0)
\label{PAS-coo-dless-semi}
\end{gather}
with the corresponding boundary condition at $\xi=0$
\begin{equation}
u'(0,\tau)=
\frac{
M\ddot u(0,\tau)+K(\epsilon\tau)u(0,\tau)-p(\tau)
}{2}.
\label{PAS-force-spr-semi}
\end{equation}
The formulation 
\eqref{PAS-coo-dless-semi},
\eqref{PAS-force-spr-semi},
\eqref{PAS-<0-SPRING}
is equivalent to the initial one 
\eqref{PAS-maineq-SPRING},
\eqref{PAS-force-spr},
\eqref{PAS-<0-SPRING}
and is used for numerical calculations
(see Section~\ref{PAS-section-numerics}).

Note that dropping out Eq.~\eqref{PAS-maineq-SPRING} and 
putting the left-hand side of Eq.~\eqref{PAS-force-spr}
to zero yields the
equation describing a forced oscillation and passage through resonance for a single degree of
freedom system \cite{kevorkian1971passage}.

\section{Free localized oscillation (some known results)}
\label{PAS-SEC:known}

Put $p=0$, $K=\const$ and consider the steady-state problem concerning the natural
oscillations of
the system described by Eqs.~\eqref{PAS-maineq-SPRING}--\eqref{PAS-force-spr}.
Take
\begin{equation}
u=\hat u(\xi)\,\exp(-i\Omega \tau).
\label{umega-SPRING}
\end{equation}
Let us show
that 
such a system possesses a mixed spectrum of natural frequencies.
There exists a continuous
spectrum of frequencies, which lies higher than the cut-off (or boundary) frequency:
$|\Omega|\geq\Omega_\ast\equiv1$.
The modes corresponding to the frequencies from the continuous spectrum are
harmonic waves. Trapped modes correspond to the frequencies from the
discrete part of the spectrum, which lies lower than the cut-off frequency:
\begin{equation}
0<|\Omega|<1.
\end{equation}
These are modes with finite energy, therefore, we require
\begin{equation}
\Int\hat u^2\,\d\xi<\infty,\qquad
\Int\hat u'{}^2\,\d\xi<\infty.
\label{finite-energy}
\end{equation}
Now we substitute Eq.~\eqref{umega-SPRING} into Eq.~\eqref{PAS-maineq-SPRING}.
This yields
\begin{equation}
 {\hat u}''
 -
(1-\Omega^2)
 {\hat u}=
 (K-M\Omega^2)\hat u(0)
 \,\delta(\xi).
 \label{PAS-ino-SPRING}
\end{equation}
The solution of Eq.~\eqref{PAS-ino-SPRING}, which satisfies 
\eqref{finite-energy}, is 
\begin{equation}
\hat u=-{(K-M\Omega^2)\hat
u(0)}\frac{\exp \big(-\sqrt{1-\Omega^2}|\xi|\big)}{2\sqrt{1-\Omega^2}}.
\label{PAS-B-SPRING-fix}
\end{equation}
Calculating the right-hand side of Eq.~\eqref{PAS-B-SPRING-fix} at $\xi=0$ yields the
equation for the trapped mode frequency:
\begin{gather}
 2\sqrt{1-\Omega_0^2} =M\Omega_0^2-K.
 \label{PAS-OSC-Omega_0^2-SPRING}
\end{gather}
Provided that restriction
\begin{equation}
-2<K<M
\label{PAS-domain}
\end{equation}
is true, there exists 
a unique trapped mode 
(see \cite{gavrilov-da70,Gavrilov_2019,gavrilov2017trapped,Glushkov2011a}).
The corresponding squared frequency $\Omega_0^2$ is
\begin{equation}
\Omega_0^2=\frac 2{M^2}\left(
\sqrt{M^2-MK+1}+\frac{MK}2-1
\right),
\label{PAS-fr-root}
\end{equation}
where $M\neq0$.
In the special case $M=0$ (an inertialess discrete oscillator, i.e. a spring
with a negative stiffness, see Eq.~\eqref{PAS-domain}), one gets
\begin{gather}
\Omega_0^2=1-\frac{K^2}4.
\end{gather}
{Inside the interval \eqref{PAS-domain}, one has}
\begin{equation}
\Omega_0{}'_K>0.
\label{PAS-Omega't}
\end{equation}
The boundary limiting values of $\Omega_0$ are
\begin{equation}
\lim_{K=-2+0}\Omega_0(K)=+0,\qquad \lim_{K=M-0}\Omega_0(K)=1-0.
\label{PAS-Omega-limits}
\end{equation}

The free localized oscillation in the case $K=K(\epsilon\tau)$ is considered in \cite{gavrilov-da70,gavrilov2017trapped}. It is shown
that the amplitude of the free localized oscillation is proportional to the
following quantity
\begin{equation}
\mathscr A
=
\frac{C_0\left( 1-{\Omega}^2_0\right)^{1/4}}
{\Omega_0^{1/2}\Big(1+M\SQRTO\Big)^{1/2}},
\label{mathscrA}
\end{equation}
where $C_0$ is a constant.
\begin{remark}	
In the limiting case 
\begin{equation}
1\ll K<M
\label{PAS-limiting-case}
\end{equation}
Eq.~\eqref{mathscrA} transforms into classical formula
\begin{equation}
\mathscr A
=
\frac{C_1}{\Omega_0^{1/2}}
\end{equation}
for a single degree of freedom system \cite{gavrilov-da70},
where 
$\Omega_0^2=K(\epsilon\tau)/M$, $C_1=C_0/M$. 
Thus, the greater the mass $M$, the closer the distributed system is to a
discrete mass-spring oscillator. The opposite (most different) limiting case is
$M=0$, $K<0$.
\label{PAS-remark1dof1}
\end{remark}

\section{Passage through resonance}
\label{PAS-SEC:pas}
The aim of this paper is to use 
the method of multiple scales 
to obtain a singular principal term of the asymptotic
expansion of $u(0,t)$ in a resonant case
\cite{kevorkian1971passage,feschenko1967eng}. In terminology of
\cite{kevorkian1971passage} we look for the principal term of the inner
expansion. 
We take a finite time interval
$[0,\tau_\ast]$, and the frequency of external excitation $\hat\Omega$ such that 
$0<\hat\Omega<1$. 

\subsection{The case of an increasing stiffness}
\label{PAS-SS1}
We assume 
(see Eqs.~\eqref{PAS-Omega't}, \eqref{PAS-Omega-limits})
that $K(\epsilon\tau)$ is 
a smooth function of $\epsilon\tau$ uniformly bounded with all its derivatives
such that
\begin{equation}
\begin{gathered}	
\dot K>0,
\\
\Omega_0\big(K(0)\big)>0,\quad
\Omega_0\big(K(
\epsilon\tau_0)\big)=\hat\Omega,\quad 
\Omega_0\big(K(
\epsilon\tau_\ast)\big)<1,
\end{gathered}
\label{PAS-Omega(K)}
\end{equation}
where $\tau_0\in[0,\tau_\ast]$ is
the instant of resonance.
We introduce slowly varying variables
\begin{equation}
T=\sqrt\epsilon(\tau-{\tau}_0),\qquad
X=\sqrt\epsilon\xi.
\label{PAS-Tdef}
\end{equation}
We represent the spring stiffness in the form of the following expansion
\begin{equation}
K(
\tau)=K_0+\epsilon K_1 (\tau-\tau_0)+o(\epsilon),
\end{equation}
or, equivalently,
\begin{equation}
K(T)=K_0+\sqrt\epsilon K_1 T+o(\sqrt\epsilon).
\label{PAS_K_expansion}
\end{equation}
Accordingly to Eqs.~\eqref{PAS-fr-root}, \eqref{PAS-Omega(K)},
$\Omega_0(K)$ is a smooth function in interval $[K(0),K(\epsilon\tau_\ast)]$, thus,
\begin{equation}
\Omega_0(\tau)=\hat{\Omega}+\epsilon {\aaa} (\tau-\tau_0)+o(\epsilon),
\label{PAS-omega-tau-expansion}
\end{equation}
or
\begin{equation}
\Omega_0(T)=\hat{\Omega}+\sqrt\epsilon {\aaa}T+o(\sqrt\epsilon).
\label{PAS-omega-expansion}
\end{equation}
Under the accepted assumptions, it follows from 
Eq.~\eqref{PAS-omega-expansion}
that
\begin{equation}
{\Omega_0'}_T=\aaa \sqrt\epsilon+o(\sqrt\epsilon),
\end{equation}
where $\aaa>0$ due to Eqs.~\eqref{PAS-Omega't}, 
\eqref{PAS-Omega(K)}.

We represent the solution in the form of the following ansatz:
\begin{gather}
 u(\xi,\tau)=W(X,T)\,\exp\varphi(\xi,\tau),
 \label{PAS-slow-repr-SPRING}
 \\
 {\varphi}'=\I\,\omega(X,T),\quad
 \dot{\varphi}=-\I\,\Omega(X,T),
 \label{PAS-fast-phases-SPRING}\\
 W(X,T)=\frac{W_0(X,T)}{\sqrt\epsilon}
 +W_1(X,T)+O(\sqrt\epsilon).
 \label{PAS-W-series-SPRING}
\end{gather}
The variables $X$, $T$, ${\varphi}$ are assumed to be independent. 
We use the following representations for the
differential operators:
\begin{equation}
\begin{gathered}
 \dot{(\cdot)}=-
 \I\,\Omega \,\partial_{\varphi }+\sqrt\epsilon\,\partial_{T}
 +O(\epsilon),
 \\
 \ddot{(\cdot)}=-\Omega ^2\,\partial^2_{\varphi \varphi }
 -2\I\sqrt\epsilon\,\Omega \,\partial^2_{\varphi T}
 -\I\sqrt\epsilon\, {\Omega '}_T\,\partial_{\varphi }
 +O(\epsilon),
 \\
 (\cdot)'=
 \I\,\omega\,\partial_{\varphi}+\sqrt\epsilon\,\partial_{X}
 +O(\epsilon),
 \\
 (\cdot)''=-\omega^2\,\partial^2_{\varphi\varphi}
 +2\I\sqrt\epsilon\,\omega\,\partial^2_{\varphi X}
 +\I\sqrt\epsilon\, {\omega}'_X\,\partial_{\varphi}
 +O(\epsilon).
\end{gathered}
\label{PAS-diff-operators}
\end{equation}
Following  
\cite{gavrilov-da70,gavrilov2002etm,indeitsev2016evolution,Gavrilov_2019}, we require that wavenumber $\omega(X,T)$ and frequency
$\Omega(X,T)$ satisfy the dispersion relation
\begin{equation}
 \omega^2
 -\Omega^2+1=0,
 \label{PAS-dis_relation-SPRING}
\end{equation}
and the equation 
\begin{equation}
{\Omega}'_X+{\omega}'_T=0,
\label{PAS-dxx-SPRING}
\end{equation}
which follows from Eq.~\eqref{PAS-fast-phases-SPRING}.
We assume that
\begin{gather}
\Omega(\pm 0,T)=\hat\Omega.
\label{PAS-Omega-Omega}
\end{gather}
Additionally, we require that 
\begin{gather}
[W]=0,\qquad 
[\varphi]=0.
\end{gather}
The phase $\varphi(\xi,\tau)$ should be defined by the formula 
\begin{equation}
 \varphi=\I\int(\omega\,\d\xi-\Omega\,\d\tau).
 \label{PAS-phase}
\end{equation}

Applying differential operators \eqref{PAS-diff-operators} to $u(\xi,\tau)$,
given by Eq.~\eqref{PAS-slow-repr-SPRING}, one obtains:
\begin{equation}
\begin{aligned}
&u'=\left(\I\hat{\omega} W+\sqrt\epsilon\frac{\partial W}{\partial X}\right)\exp
(-\I\hat{\Omega}\tau)
 +O(\epsilon)
,\\
&u''=\left(
-{\hat{\omega}}^2 W
+2\I \sqrt\epsilon\hat{\omega}\frac{\partial W}{\partial X}
\right)\exp(-\I\hat{\Omega}\tau)
 +O(\epsilon)
,\\
&\ddot{u}=\left( -\hat\Omega^2 W
-2\I \sqrt\epsilon\hat{\Omega}\frac{\partial W}{\partial T} 
\right)\exp(-\I\hat{\Omega}\tau)
 +O(\epsilon).
\end{aligned}
\end{equation}
Here the wavenumber $\hat{\omega}=\pm\I\sqrt{1-{\hat{\Omega}}^2}$ corresponds 
to the frequency $\hat\Omega$ due to 
dispersion
relation \eqref{PAS-dis_relation-SPRING}.
Taking into account Eqs.~\eqref{PAS-W-series-SPRING}, 
we get:
\begin{gather}
\begin{aligned}	
&u'=\Bigg( 
\frac{\I\hat{\omega}\,W_0}{\sqrt\epsilon}
+
\bigg(
\frac{\partial W_0}{\partial X} 
+\I\hat{\omega}W_1      \bigg)
\Bigg)
\exp(-\I\hat{\Omega}\tau)
+O(\sqrt\epsilon),
\\
&u''=\Bigg(
-\frac{\hat{\omega}^2 W_0}{\sqrt\epsilon}
+\left(
2\I\hat{\omega}\frac{\partial W_0}{\partial X}
-\hat{\omega}^2 W_1
\right)
\Bigg)
\exp(-\I\hat{\Omega}\tau)+O(\sqrt\epsilon),
\\
&\ddot{u}=
\Bigg(
-\frac{\hat{\Omega}^2 W_0}{\sqrt\epsilon}
+
\bigg(
-2\I\hat{\Omega}\frac{\partial W_0}{\partial T}
-\hat{\Omega}^2 W_1
\bigg)
\Bigg)
\exp(-\I\hat{\Omega}\tau)+O(\sqrt\epsilon).
\end{aligned}
\end{gather}
Substituting the above expressions into Eq.~\eqref{PAS-1dof-SPRING-p} and
equating coefficients of like powers $\epsilon$, one obtains that the term
of order $\epsilon^{-1/2}$ equals zero identically due to frequency
equation~\eqref{PAS-OSC-Omega_0^2-SPRING}. The zeroth order term is
\begin{equation}
\left[\frac{\partial W_0}{\partial X}  \right]
=
K_1 T W_0
-1
-
2\I M\hat{\Omega}\frac{\partial W_0}{\partial T}.
\label{PAS-the-jump}
\end{equation}
Note that putting the left-hand side of the last equation to zero yields the
equation describing passage through the resonance for a single degree of
freedom system \cite{kevorkian1971passage}.

The unknown quantity $\left[\frac{\partial W_0}{\partial X}  \right]$ 
in the left-hand side of Eq.~\eqref{PAS-the-jump}
can be defined
by consideration of 
Eq.~\eqref{PAS-maineq-SPRING} at $\xi=\pm0$. To do this,
we substitute ansatz 
\eqref{PAS-slow-repr-SPRING}--\eqref{PAS-W-series-SPRING}
and representations \eqref{PAS-diff-operators}
into Eq.~\eqref{PAS-maineq-SPRING} 
and equate coefficients of like powers $\epsilon$. 
Taking into account dispersion relation 
\eqref{PAS-dis_relation-SPRING},
one obtains
\begin{equation}
\left[\frac{\partial W_0}{\partial X}  \right]
=\frac{2\I\hat{\Omega}}{\sqrt{1-\hat{\Omega}^2}}\frac{\partial W_0}{\partial
T}.
\label{PAS-via-gov-equation}
\end{equation}
Now we equate the right-hand sides of 
Eqs.~\eqref{PAS-the-jump}, \eqref{PAS-via-gov-equation}, and get the
equation for $\W(T)\equiv W_0(0,T)$:
\begin{equation}
-\I\frac{\partial \W}{\partial T}+\frac{1}{2}K_1f(\hat{\Omega})T \W=
\frac{f(\hat{\Omega})}{2}.
\label{PAS-W0-equation}
\end{equation}
Here
\begin{equation}
f(\hat{\Omega})=
\frac{\sqrt{1-\hat{\Omega}^2}}{\hat{\Omega}\big(1+M\sqrt{1-\hat{\Omega}^2}\big)}.
\label{PAS-def-f}
\end{equation}
Substituting expressions~\eqref{PAS_K_expansion}, \eqref{PAS-omega-expansion}
into frequency equation~\eqref{PAS-OSC-Omega_0^2-SPRING} and equating coefficients
of like powers, one can demonstrate that
\begin{equation}
\aaa=\frac{1}{2}K_1 f(\hat{\Omega}).
\label{PAS-omega_01-proportional-to-K1}
\end{equation}
Hence, Eq.~\eqref{PAS-W0-equation} can be written as follows:
\begin{equation}
-\I\frac{\partial \W}{\partial T}+\aaa T \W=
\frac{f(\hat{\Omega})}{2}.
\label{PAS-W0-equation-1}
\end{equation}
The case of
passage through the resonance for a single degree of
freedom system \cite{kevorkian1971passage} can be formally obtained by the choice
\begin{equation}
f(\hat{\Omega})=f_0(\hat\Omega)\equiv
\frac{1}{M\hat{\Omega}}.
\label{PAS-def-f-single}
\end{equation}
\begin{remark}	
In the limiting case 
\eqref{PAS-limiting-case},
one gets 
\begin{equation}
f(\hat\Omega)\simeq f_0(\hat\Omega)
\end{equation}
(see also Section~\ref{PAS-SEC:known} and \cite{gavrilov-da70}). Therefore, in the 
case \eqref{PAS-limiting-case}, the principal term of the asymptotic
solution for the problem under
consideration transforms into the principal term of the solution for the
corresponding problem for a linear oscillator. 
Thus, the greater the mass $M$, the closer the distributed system is to a
discrete mass-spring oscillator.
The opposite (most different)
limiting case is
$M=0$, $K<0$.
The same conclusions are true for a free
localized oscillation in the system under consideration 
(see Remark~\ref{PAS-remark1dof1}).
\label{PAS-remark1dof}
\end{remark}

According to 
Eqs.~\eqref{PAS-Omega(K)},
\eqref{PAS-omega_01-proportional-to-K1}, one has $\aaa>0$.
To solve Eq.~\eqref{PAS-W0-equation-1}, following to 
\cite{kevorkian1971passage}, 
and taking into account the last inequality,
we introduce the new variable~$\eta$
\begin{equation}
\eta=\frac{|\aaa|}{2}T^2.
\label{PAS-eta-ansatz}
\end{equation}
We can now rewrite Eq.~\eqref{PAS-W0-equation-1} as follows:
\begin{equation}
\frac{\partial \W^{\pm}}{\partial \eta}+\I \W^{\pm}=\pm \frac{\I f(\hat{\Omega})}
{2\sqrt{2 |\aaa|\eta}}.
\label{PAS-W0-equation-eta}
\end{equation}
Here and in what follows superscript ``$-$'' corresponds to the case $T<0$, and 
``$+$'' corresponds to $T>0$.
We search a solution of Eq.~\eqref{PAS-W0-equation-eta} in the following form:
\begin{equation}
\W^{\pm}=a^{\pm}\exp{(-\I \eta)}+P^{\pm}(\eta),
\end{equation}
where $a^{\pm}$ are unknown constants, 
\begin{equation}
P^{\pm}(\eta)=
\mp\frac{\I f(\hat{\Omega})}{2\sqrt{2 |\aaa|}}
\int_{\eta}^{\infty}\frac{\exp{\big(-\I(\eta-s)\big)}}{\sqrt s}\,\d s.
\label{PAS_P(eta)}
\end{equation}
One has
\begin{multline}
\Phi(\eta)\equiv
\int_{\eta}^{\infty}\frac{\exp{\big(-\I(\eta-s)\big)}}{\sqrt s}\,\d s
\\
=
-\sqrt{\frac\pi2}(1+\I)\left(\erf\left(\sqrt\frac\eta2(1-\I)\right)-1\right)\mathrm
e^{-\I\eta},
\end{multline}
\begin{gather}
\Re\Phi(\eta)
=
\sqrt{2\pi}\Bigg(
-\cos\eta \,\,C\! \left(\sqrt{\frac{2\eta}\pi}\right)
-\sin\eta \,\,S\! \left(\sqrt{\frac{2\eta}\pi}\right)
+\frac{\cos\eta}2
+\frac{\sin\eta}2
\Bigg),
\\
\Im\Phi(\eta)
=
\sqrt{2\pi}\Bigg(
\sin\eta \,\,C\! \left(\sqrt{\frac{2\eta}\pi}\right)
-\cos\eta \,\,S\! \left(\sqrt{\frac{2\eta}\pi}\right)
+\frac{\cos\eta}2
-\frac{\sin\eta}2
\Bigg),
\\
\Phi(0)=\frac{\sqrt{2\pi}(1+\I)}{2},
\end{gather}
where $\erf(\cdot)$ is the error function \cite{abramowitz1972handbook},
$C(\cdot)$, $S(\cdot)$ are the normalized
Fresnel integrals \cite{abramowitz1972handbook}.

One can put \cite{kevorkian1974erratum}
\begin{equation}
a^{-}=0.
\end{equation}
Since the solution must be continuous at the instant $T=0$, we require
\begin{equation}
a^{+}+P^{+}(0)=P^{-}(0).
\end{equation}
By virtue of Eq.~\eqref{PAS_P(eta)}, one gets:
\begin{equation}
a^{+}=\frac{\I f(\hat{\Omega})}{\sqrt{2 |\aaa|}}\,
\Phi(0).
\end{equation}
Hence,
\begin{equation}
a^{+}=\frac{(-1+\I)\sqrt{\pi} f(\hat{\Omega})}{2\sqrt{|\aaa|}}.
\end{equation}
Thus, using Eq.~\eqref{PAS-eta-ansatz} for  $\eta$,  we get:
\begin{gather}
\W^{-}=\frac{\I f(\hat{\Omega})}{2\sqrt{2 |\aaa|}}
\,\Phi\left(\frac{|\aaa| T^2}{2}\right)
,
\label{PAS-W0-}
\\
\W^{+}=\frac{(-1+\I)\sqrt{\pi} f(\hat{\Omega})}{2\sqrt{|\aaa|}}
\exp\left(-\frac{\I |\aaa| T^2}{2}\right)
-
\frac{\I f(\hat{\Omega})}{2\sqrt{2 |\aaa|}}
\,\Phi\left(\frac{|\aaa| T^2}{2}\right)
,
\label{PAS-W0+}
\\
u(0,\tau)=\frac{\W^{-}H(-T)+\W^{+}H(T)}{\sqrt\epsilon}\exp(-\I\hat\Omega
\tau)+O(1),
\label{PAS-final}
\end{gather}
where $f(\hat\Omega)$ is defined by Eq.~\eqref{PAS-def-f}. 
Note that taking here $f(\hat\Omega)$ in the form of 
Eq.~\eqref{PAS-def-f-single} yields the classical result for a single degree of
freedom system 
\cite{kevorkian1971passage,kevorkian1974erratum}.
To obtain the physical displacements that correspond to the real part of the
right-hand side of Eq.~\eqref{PAS-external-force}, one need to take the real part 
of Eq.~\eqref{PAS-final}.

\subsection{The case of a decreasing stiffness}
\label{PAS-SS2}
Consider now the case, when
\begin{equation}
\begin{gathered}	
\dot K<0,
\\
\Omega_0\big(K(0)\big)<1,\quad
\Omega_0\big(K(\epsilon\tau_0)\big)=\hat\Omega,\quad
\Omega_0\big(K(\epsilon\tau_\ast)\big)>0.
\end{gathered}
\label{PAS-Omega(K)-1}
\end{equation}
In this case, the quantity $K_1$ in Eq.~\eqref{PAS_K_expansion} is such that
\begin{equation}
K_1<0.
\end{equation}
Hence,  due to Eq.~\eqref{PAS-omega_01-proportional-to-K1}, the quantity
$\Omega_{01}$ in Eq.~\eqref{PAS-omega-expansion} is also negative:
\begin{equation}
\Omega_{01}<0.
\end{equation}
Taking into account the last inequality, and substituting Eq.~\eqref{PAS-eta-ansatz} into
Eq.~\eqref{PAS-W0-equation-1}, analogously to the case $K_1>0$ one gets
\begin{equation}
u(0,\tau)=-\frac{\overline{\W^{-}}H(-T)+
\overline{\W^{+}}H(T)}{\sqrt\epsilon}\exp(-\I\hat\Omega\tau)+O(1).
\label{PAS-back-pass}
\end{equation}
Here $\overline{\W^{-}}$,  $\overline{\W^{+}}$ denote complex
conjugates of functions $\W^{-}$, $\W^{+}$ (given by Eqs.~\eqref{PAS-W0-},
and \eqref{PAS-W0+}), respectively.

\section{Numerics}
\label{PAS-section-numerics}

Unlike the case of a single degree of freedom system, the applicability of the
suggested asymptotic approach to the non-stationary problem for the
considered distributed system can be verified only numerically.
One of the reasons for this is the presence of contributions of all 
frequencies in a non-stationary solution.
Such a verification is 
the main aim of calculations presented in this section.
%
For numerical study we use 
{\sc SciPy} software.
The discretization of PDE 
\eqref{PAS-coo-dless-semi}
is defined in accordance with the following scheme:
\begin{equation}
\frac{u_{j+1}^{i}-2u_j^{i}+u_{j-1}^{i}}{(\Delta \xi)^2}-
\frac{u_j^{i+1}-2u_j^{i}+u_j^{i-1}}{(\Delta \tau)^2}-
\frac{u_j^{i+1}+u_j^{i-1}}2=0,
\end{equation}
where integers $i,\ j$ ($0\leq j\leq N,\ 1\leq i$) are such that
\begin{gather}
u_j^i=u(j\Delta\xi,i\Delta\tau).
\end{gather}
This scheme conserves 
\cite{donninger2011numerical,strauss1978numerical}
the discrete energy for a
nonlinear Klein-Gordon equation with constant coefficients. 
Numeric boundary conditions that correspond to
\eqref{PAS-force-spr-semi} are taken in the form \cite{strikwerda2004finite}
\begin{multline}
\frac{-3u_0^{i+1}+4u_1^{i+1}-u_2^{i+1}}{2\Delta\xi}
+
\frac{-3u_0^{i-1}+4u_1^{i-1}-u_2^{i-1}}{2\Delta\xi}
\\-
\frac{K^{i+1}u_0^{i+1}+K^{i-1}u_0^{i-1}}2
-
M\,\frac{u_0^{i+1}-2u_0^{i}+u_0^{i-1}}{(\Delta \tau)^2}
+
\frac{p^{i+1}+p^{i-1}}2=0,
\end{multline}
where $i>1$, 
\begin{equation}
K^i=K(i\Delta\tau).
\end{equation}
At the right end of the string, we use the condition
\begin{equation}
u^i_{N}=u^i_{N-1}\qquad (i>1)
\end{equation}
\NEW{that corresponds to the physical boundary condition
$u'=0$. Actually, the specific form of this boundary condition
is not very important in our calculations, since we consider the discrete model of
the string with sufficiently large length such that the wave reflections at the right
end do not occur.}
Numerical initial conditions are 
\begin{equation}
u^0_j=u^{-1}_j=0.
\end{equation}
All numerical results in what follows are obtained for the choice
\begin{equation}
\Delta\xi=0.008,\qquad 
\Delta\tau=0.002. 
\label{Deltas}
\end{equation}

\begin{figure}[hp]
\centering{\includegraphics[width=0.98\textwidth]{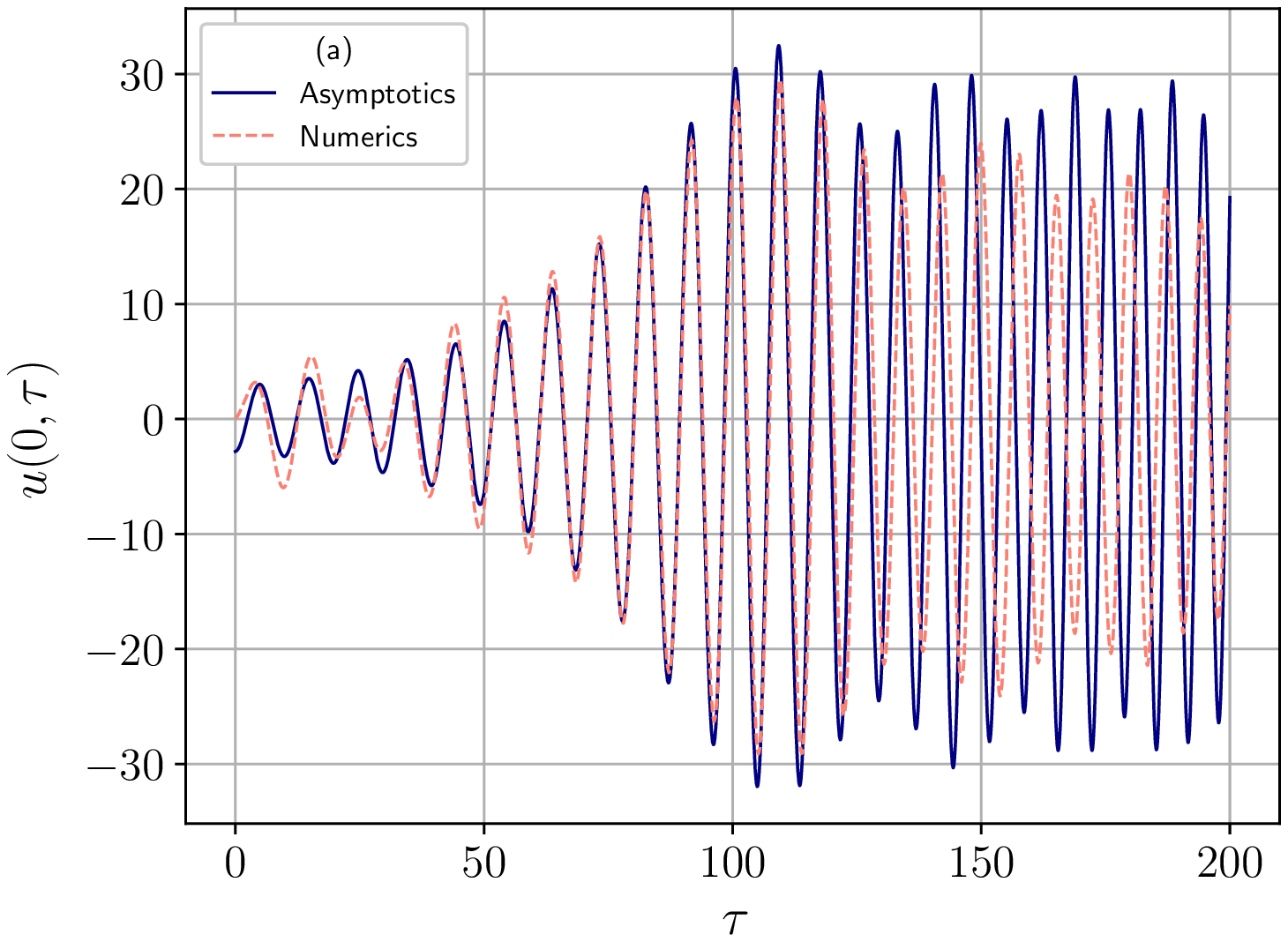}
\includegraphics[width=0.98\textwidth]{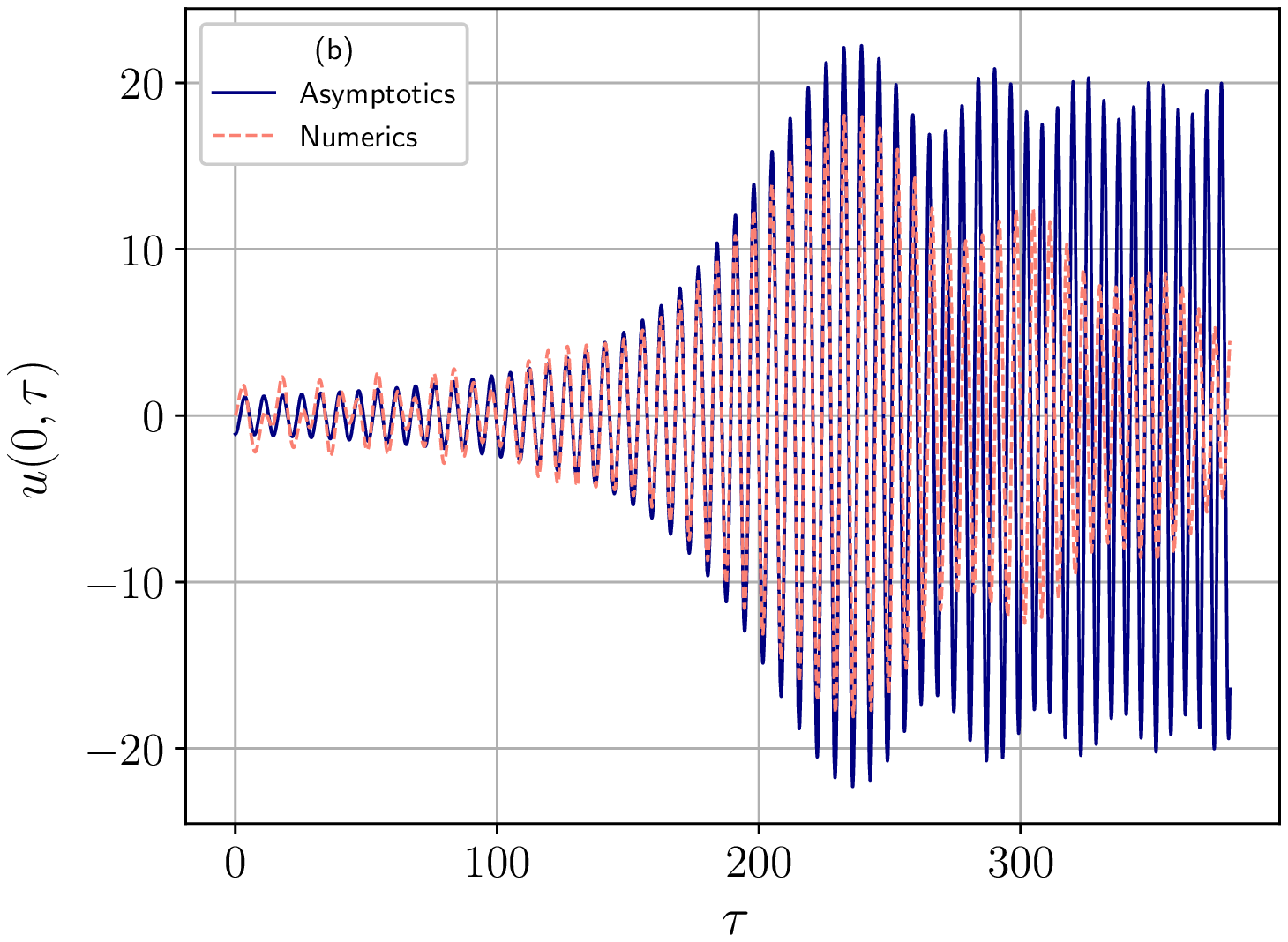}
}
\caption{
Comparing the asymptotic solution in the form of 
Eqs.~\eqref{PAS-def-f},
\eqref{PAS-W0-},
\eqref{PAS-W0+},
\eqref{PAS-final}
and the numerical solution for inertialess oscillator with an increasing stiffness
$K(\epsilon\tau)=-1.9+0.005\tau$,
(a) $\hat\Omega^2=0.4$, the instant of resonance $\tau_0\simeq70.16$,
$K(\epsilon\tau_0)\simeq-1.55$,
(b) $\hat\Omega^2=0.75$, $\tau_0=180.0, K(\epsilon\tau_0)=-1.0$ }
\label{passage04.pdf}
\end{figure}

\begin{figure}[hp]
\centering{\includegraphics[width=0.95\textwidth]{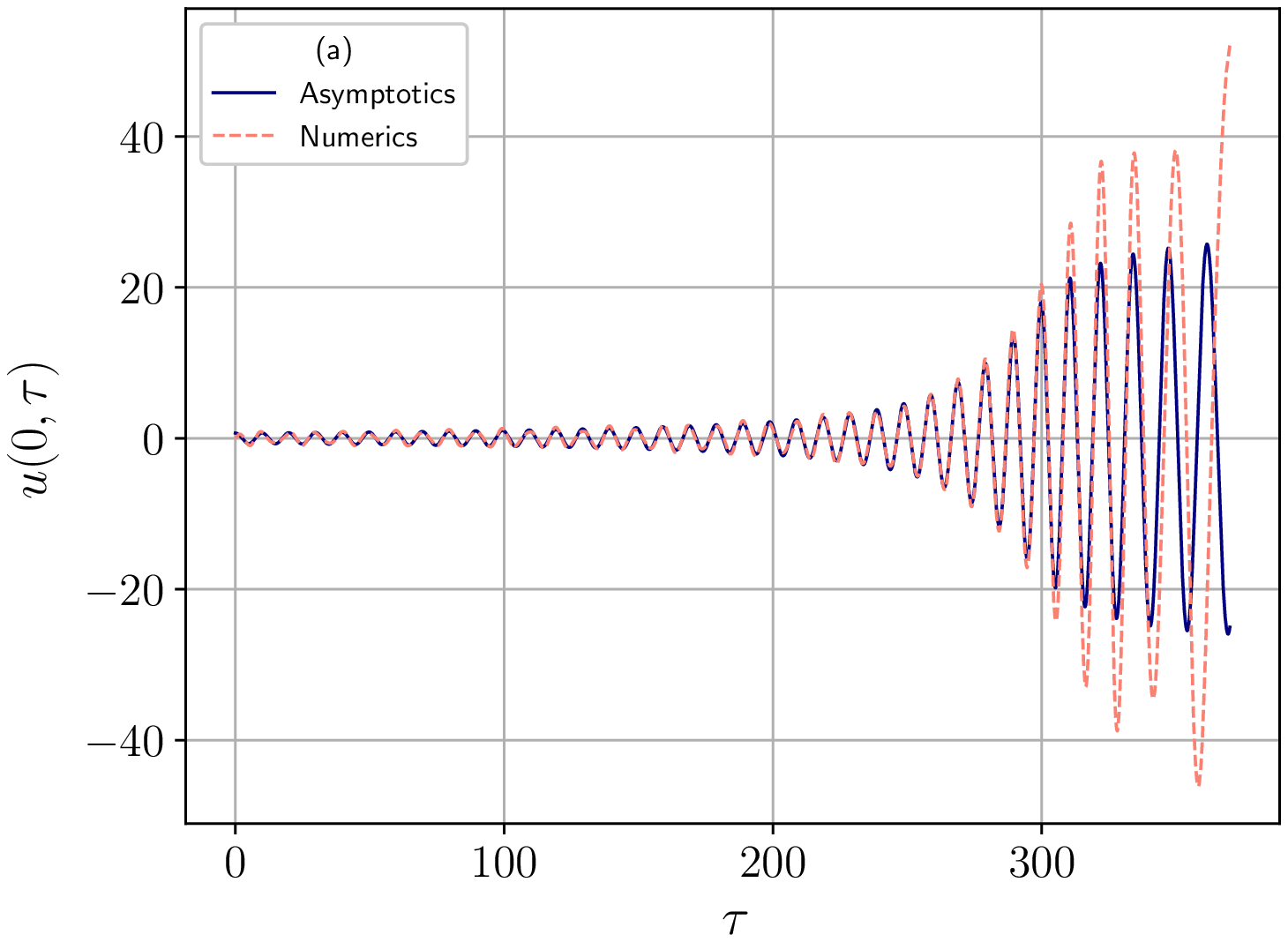}
\includegraphics[width=0.95\textwidth]{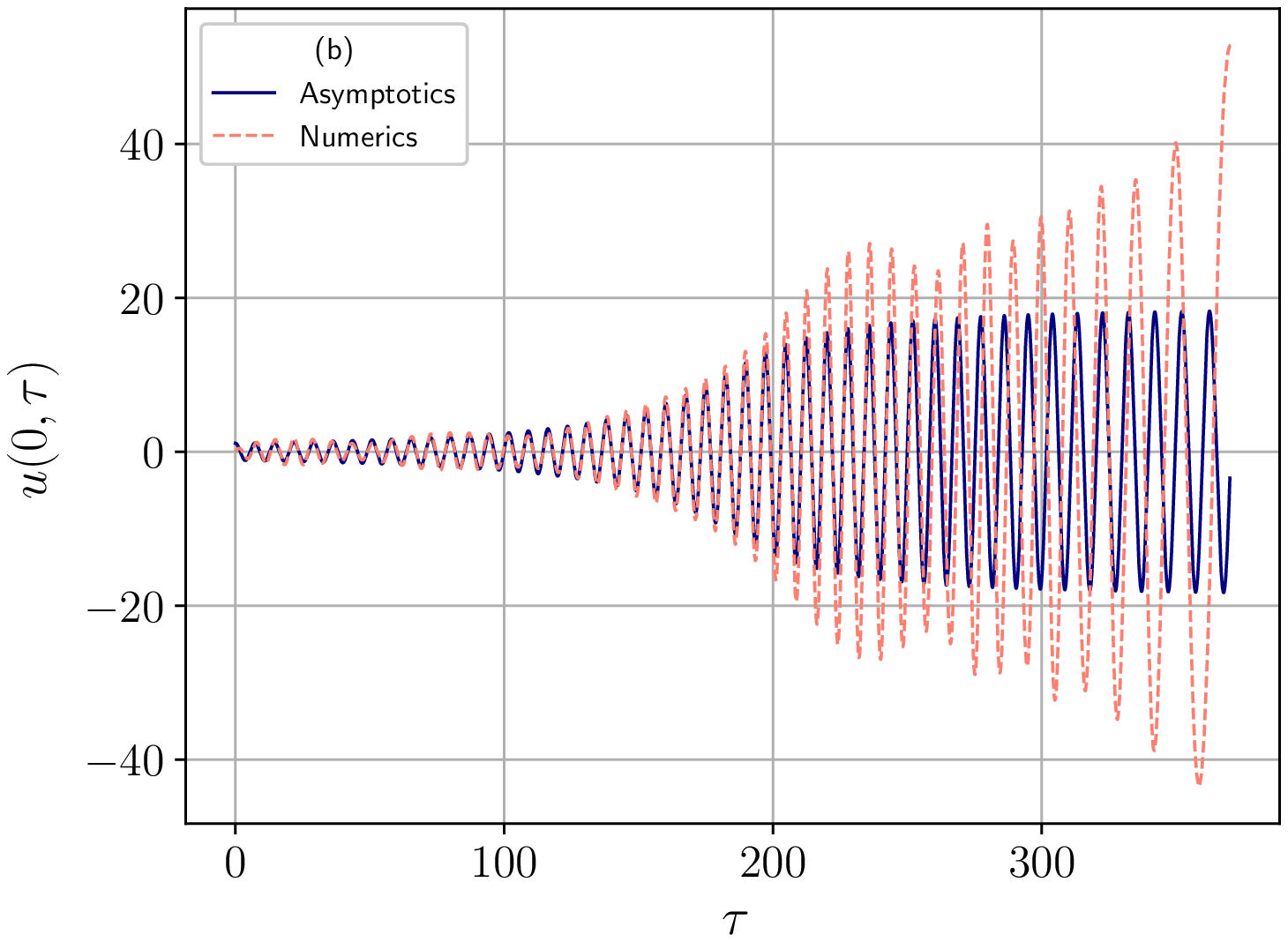}
}
\caption{
Comparing the asymptotic solution in the form of 
Eqs.~\eqref{PAS-def-f},
\eqref{PAS-W0-},
\eqref{PAS-W0+},
\eqref{PAS-back-pass}
and the numerical solution for inertialess oscillator with a decreasing
stiffness
$K(\epsilon\tau)=-0.1-0.005\tau$.
(a) $\hat\Omega^2=0.4$, the instant of resonance 
$\tau_0\simeq289.84,\ K(\epsilon\tau_0)\simeq-1.55$,
(b) $\hat\Omega^2=0.75$, $\tau_0=180.0,\ K(\epsilon\tau_0)=-1.0$
}
\label{back04.pdf}
\end{figure}

\begin{figure}[hp]
\centering{\includegraphics[width=0.95\textwidth]{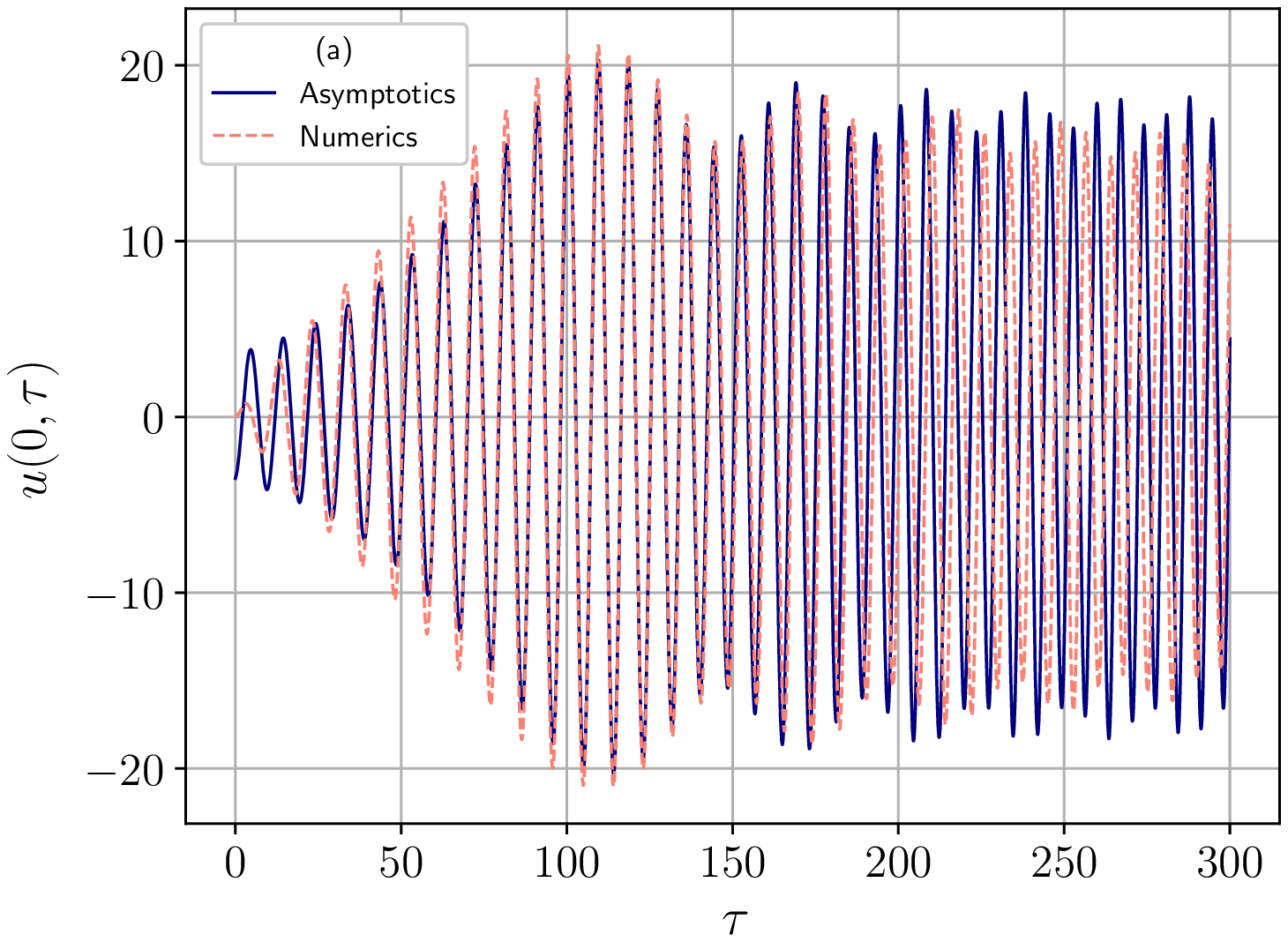}
\includegraphics[width=0.95\textwidth]{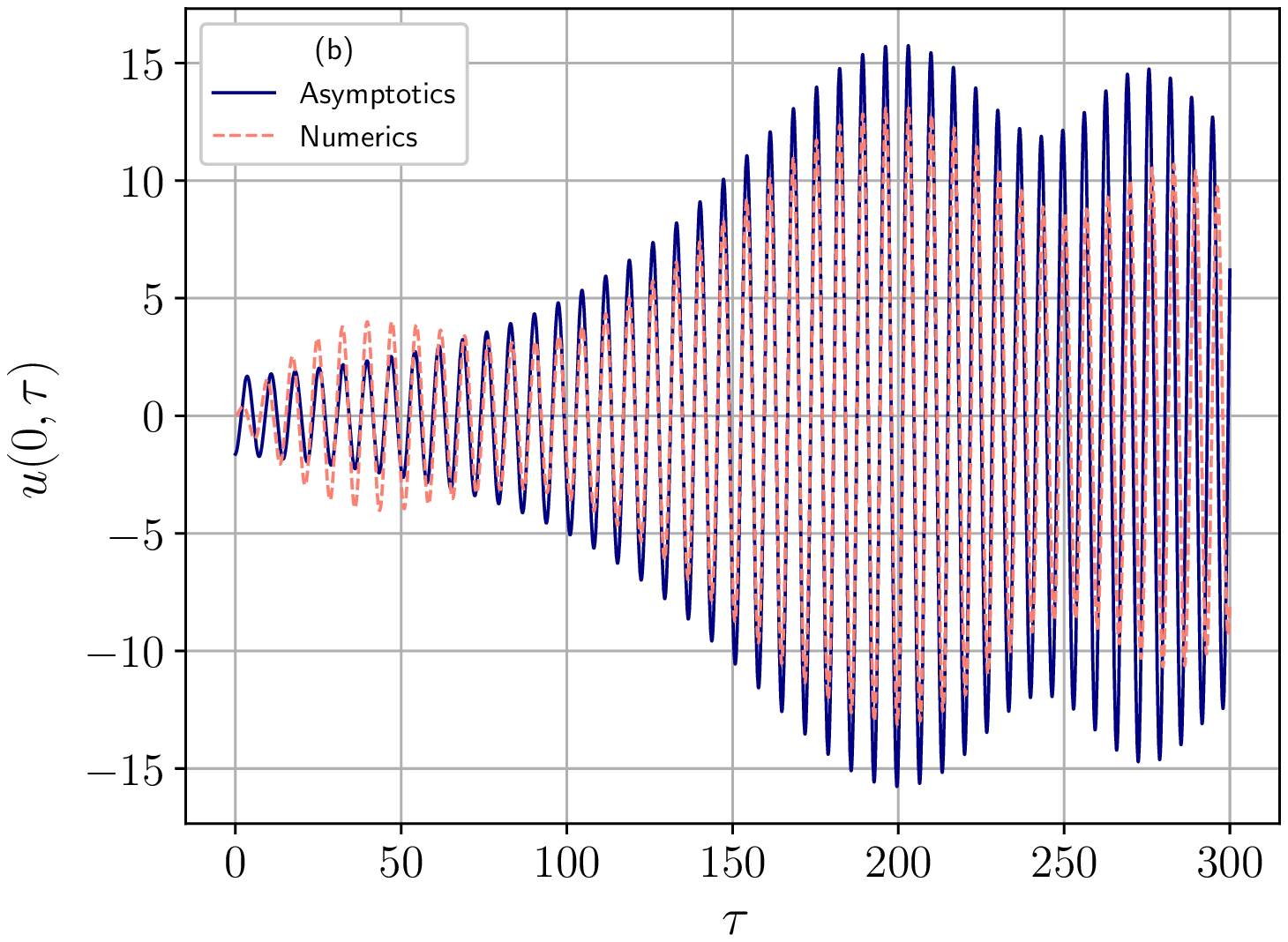}
}
\caption{
Comparing the asymptotic solution in the form of 
Eqs.~\eqref{PAS-def-f},
\eqref{PAS-W0-},
\eqref{PAS-W0+},
\eqref{PAS-final}
and the numerical solution for massive oscillator ($M=2$) with an increasing stiffness.
(a) $\hat\Omega^2=0.4$, 
$K(\epsilon\tau)=-1.0+0.005\tau$,
the instant of resonance $\tau_0\simeq50.16$,
$K(\epsilon\tau_0)\simeq-0.75$,
(b) $\hat\Omega^2=0.75$, 
$K(\epsilon\tau)=-0.1+0.005\tau$,
$\tau_0=120.0$,
$K(\epsilon\tau_0)=0.5$
}
\label{M2passage04.pdf}
\end{figure}

In Figure~\ref{passage04.pdf} we compare
the analytical and numerical results in the case of inertialess oscillator 
$M=0$ with an increasing stiffness (see Remark~\ref{PAS-remark1dof}, which
clarifies the motivation of such a choice).
In Figure~\ref{back04.pdf} we compare
the results for the same system 
in the case of a decreasing stiffness. In 
Figure~\ref{M2passage04.pdf} we compare the results for the system with a
massive oscillator, where $M=2$ in the case of an increasing stiffness.

One can see that the comparison demonstrates a good mutual agreement in a
neighbourhood of the instant of resonance, and suggested asymptotic approach 
is really applicable to the problem under consideration.
One can observe also that the influence
of the term of order $O(1)$ is more noticeable in the case of a massive
oscillator.

\section{Conclusion}
\label{PAS-SEC:Concl}
The most important result of the paper is the expressions
(see Eqs.~\eqref{PAS-W0-},
\eqref{PAS-W0+},
\eqref{PAS-final},
and 
\eqref{PAS-back-pass})
for the principal singular term of order $O(1/\sqrt{\epsilon})$ of the
resonant solution describing a forced oscillation of an infinite-length
mechanical system
possessing a single trapped
mode. The system is a string, lying on the Winkler
foundation and equipped with a discrete linear mass-spring oscillator of 
time-varying stiffness,
see Eqs.~\eqref{PAS-maineq-SPRING},
\eqref{PAS-force-spr}. 
According to Eq.~\eqref{PAS-Tdef}, the diameter of the neighbourhood,
where the resonant solution is applicable, has an order $O(1/\sqrt\epsilon)$.
The results can be 
generalized for other mechanical systems possessing a single
trapped mode.

\begin{NEWe}    
We have to emphasize that the structure of the first approximation equation \eqref{PAS-the-jump} 
for the problem under
consideration is quite different comparing to the problem on passage through
resonance for a linear oscillator. The latter case corresponds
to Eq.~\eqref{PAS-the-jump}, where the left-hand side equals zero. 
Assuming additionally that we deal with the special case $M=0$ 
(discussed in Remarks~\ref{PAS-remark1dof1}, \ref{PAS-remark1dof}),
we get the first approximation 
equation\footnote{\NEW{In the form of Eq.~\eqref{PAS-the-jump}, where two
corresponding terms equal zero}}, which is,
clearly, senseless.  Thus, the presented in this paper analytical solution
cannot be obtained
straightforwardly using known solutions
\cite{kevorkian1971passage,kevorkian1974erratum}
for a linear oscillator. 
\end{NEWe}
Nevertheless, the obtained final solution has a similar structure with the corresponding
solution for a linear oscillator
\cite{kevorkian1971passage,kevorkian1974erratum}. 
The difference is in the structure of function
$f(\hat\Omega)$ (see Eqs.~\eqref{PAS-def-f}, \eqref{PAS-def-f-single}). Note
that in the limiting case 
\eqref{PAS-limiting-case} the solution transforms into the corresponding
solution for a linear oscillator, see Remark~\ref{PAS-remark1dof}. 
The obtained analytical solution was verified by independent numerical
calculations and a good agreement
in a
neighbourhood of the instant of resonance was demonstrated.

In order to obtain a uniformly valid asymptotic solution for both
non-resonant and resonant cases, we definitely need to obtain the next term
(of order $O(1)$) for the resonant solution.  This, probably, will allow
us to match the corresponding terms in the resonant and non-resonant
solutions.  Note that the problem to determine the next term of the ``inner
expansion'' seems to be much more complicated than the one considered in this
paper. On the other hand, the expression for the next term of the resonant
solution will contain unknown constants, which should be determined as a
result of matching \cite{kevorkian1971passage,kevorkian1974erratum}.
Thus, the expression for the next term cannot be verified numerically
before matching is done.

\begin{NEWe}    
Note that for the time being we do not know how to take into account properly the
possible presence of a viscous friction (i.e., a dissipation) considering
non-stationary oscillations in the system under consideration.
The presence of a dissipation makes the problem more
sophisticated. 
It may be a subject of a separate future study. 
\end{NEWe}

The results obtained in the paper can be useful, in particular, 
for investigation of the internal
resonances in a linear infinite-length system, having time-varying parameters and 
possessing several trapped modes 
with corresponding frequencies close
to each other (e.g., a string on the Winkler foundation with several distant moving
discrete inclusions, see 
\cite{Vesnitskii1992,indeitsev2000resonance}). 

\section*{Acknowledgements}
The authors are grateful to D.A.~Indeitsev for useful and stimulating discussions.

\section*{Declaration of interest}
None
\ifdefined\DIFadd
\def\href{} 
\def\path{}
\newcommand{\startunderscoreletter}{\catcode`_ 12\relax}
\startunderscoreletter
\else
\fi

\bibliographystyle{elsarticle-num}
\biboptions{compress}
\bibliography{serge-gost,in-library,mode,mode-trans,journals.rus,string-spring,math,metamat,passage}